\documentclass[conference]{IEEEtran}
\IEEEoverridecommandlockouts
\usepackage{cite}
\usepackage{amsmath,amssymb,amsfonts}
\usepackage{algorithmic}
\usepackage{graphicx}
\usepackage{textcomp}
\usepackage{xcolor}
\def\BibTeX{{\rm B\kern-.05em{\sc i\kern-.025em b}\kern-.08em
    T\kern-.1667em\lower.7ex\hbox{E}\kern-.125emX}}
\begin{document}

\title{Towards Ultra-Low Latency: Binarized Neural Network Architectures for In-Vehicle Network Intrusion Detection\\
\thanks{This study was partially funded by the budget project FFZF-2025-0016.}
}

\author{\IEEEauthorblockN{Huiyao Dong}
\IEEEauthorblockA{\textit{Faculty of Information Technology and Security} \\
\textit{ITMO University}\\
Saint Petersburg, Russia \\
hydong@itmo.ru}
\and
\IEEEauthorblockN{Igor Kotenko}
\IEEEauthorblockA{\textit{Laboratory of Computer Security Problems} \\
\textit{St. Petersburg Federal Research Center} \\ 
\textit{of the Russian Academy of Sciences (SPC RAS)}\\
Saint Petersburg, Russia \\
ivkote1@mail.ru}
}

\maketitle

\begin{abstract}
The Control Area Network (CAN) protocol is essential for in-vehicle communication, facilitating high-speed data exchange among Electronic Control Units (ECUs). However, its inherent design lacks robust security features, rendering vehicles susceptible to cyberattacks. While recent research has investigated machine learning and deep learning techniques to enhance network security, their practical applicability remains uncertain. This paper presents a lightweight intrusion detection technique based on Binarized Neural Networks (BNNs), which utilizes payload data, message IDs, and CAN message frequencies for effective intrusion detection. Additionally, we develop hybrid binary encoding techniques to integrate non-binary features, such as message IDs and frequencies. The proposed method, namely the BNN framework specifically optimized for in-vehicle intrusion detection combined with hybrid binary quantization techniques for non-payload attributes, demonstrates efficacy in both anomaly detection and multi-class network traffic classification. The system is well-suited for deployment on micro-controllers and Gateway ECUs, aligning with the real-time requirements of CAN bus safety applications.
\end{abstract}


\section{Introduction}
The CAN protocol is essential for in-vehicle communication due to its efficiency in real-time data exchange among ECUs. Designed for robust functionality, it enables high-speed communication at standard rates of up to 1 Mbps~\cite{ecu1}. The CAN FD variant further enhances this capability, supporting data rates of 10-12 Mbps~\cite{ecu2}. However, the protocol prioritizes minimal overhead and deterministic responses over strong security features such as encryption and authentication. This design introduces vulnerabilities: the lack of authentication allows any ECU to send messages; the absence of encryption makes traffic susceptible to eavesdropping and modification; and the broadcast nature means a single compromised ECU can jeopardize the entire vehicle~\cite{salfer2015attack}. As a result, vehicles are vulnerable to various cyberattacks, including message injection, DoS, and spoofing~\cite{kang2024canival}. Consequently, advancing intrusion detection systems (IDSs) has become critical.

Recent research has explored machine learning (ML) and deep learning (DL) techniques to enhance in-vehicle network security, showing promise in malicious node detection~\cite{node1}, anomaly detection~\cite{ano1}, and CAN message frequency-based intrusion detection~\cite{freq1}. While these techniques demonstrate high performance on public datasets, their practical applicability is uncertain~\cite{idssurvey}. DL models often require significant computational resources, complicating deployment and future optimization~\cite{key_02,key_06}, while lightweight ML algorithms face challenges in adapting to evolving attack vectors~\cite{simple1,simple2}. Further investigation into methods that are both effective and computationally efficient is warranted.

Binarized neural networks (BNNs)~\cite{2016binarized} are ultra-lightweight deep learning models with 1-bit weights and activations (+1 or -1), significantly enhancing computational efficiency. Their minimal memory requirements make them suitable for deployment on micro-controllers and gateway ECUs, while their ultra-low latency meets the real-time demands of CAN bus safety applications. BNNs excel at learning patterns in binary payload data, which is crucial for security analysis. Despite their advantages, a significant challenge in applying BNNs to in-vehicle intrusion detection is their inherent limitation in effectively processing the heterogeneous and often non-binary nature of CAN bus features beyond mere payload data, such as message IDs and frequencies. 

This paper proposes a novel BNN-based lightweight IDS that not only uses payload data like previous works~\cite{bnn1, bnn2, bnn3}, but also innovatively integrates message IDs and CAN message frequencies for robust intrusion detection, demonstrating effectiveness in both anomaly detection and multi-class network traffic classification. The key contributions include: (1) a BNN framework optimized for in-vehicle intrusion detection; (2) hybrid binary quantization techniques for non-payload attributes and a dual-path processing mechanism to enhance BNN compatibility with heterogeneous features; and (3) comprehensive experiments on two datasets across various tasks, benchmarking against conventional DL models to evaluate performance-complexity trade-offs using metrics such as detection rates, accuracy, model size, and response time.

\section{Related Work}
Early intrusion detection techniques were mainly based on straightforward ML models. Entropy-based detection achieves over 80\% recall for flooding attacks with near-zero false positives under stable conditions~\cite{wevj16060334}, while OCSVM~\cite{simple1} and hierarchical clustering with dynamic time warping~\cite{simple2} have demonstrated perfect success rates in detecting frequency anomalies. Although these comparatively simple techniques are well suited for legacy ECUs, they may be less effective against payload manipulation and often lack adaptability to dynamic network traffic.

The evolving dynamics of in-vehicle networks and increasingly complex tasks require the adoption of DL techniques. Although naive DNNs~\cite{okokpujie2021anomaly} have demonstrated significant improvements in detection accuracy, hybrid models combining 1D-CNNs with LSTMs effectively capture both localised payload characteristics and temporal dependencies~\cite{10556797}. Maliha et al.~\cite{maliha2024unified} proposed a 2-tier time series-based anomaly intrusion detection model that processed short-term and long-term memory separately to learn the deviation of benign stateful latent space and identify attacks deviating from benign samples.
Recent methods also increasingly incorporate self-attention mechanisms to enhance intrusion and anomaly detection performance. The integration of self-attention in hybrid detectors has been validated in autoencoder~\cite{key_01}, GRU-CNN-based architectures\cite{key_04} and LSTM-based multi-task configurations~\cite{key_05}. Others utilize vanilla transformer blocks. Le et al.\cite{LE2024112091} proposed a framework leveraging autoencoders and sequence-level transformers, achieving a 100\% detection rate, while Nguyen et al.\cite{key_02} adapted vanilla transformer blocks for dense intrusion detection. However, the high complexity and resource demands of these models limit their practicality for in-vehicle environments.

While most studies pursue enhanced performance at the expense of increased complexity, several have demonstrated the value of efficient techniques. Zhang et al.\cite{bnn1} introduced a BNN-based IDS for in-vehicle networks, utilizing binary activations and weights to accelerate intrusion detection and reduce both memory and energy consumption, achieving a significant decrease in detection latency on standard CPUs. This approach was further extended by investigating BNN design parameters~\cite{bnn2}. BCNN-based IDSs were proposed to exploit both temporal and spatial correlations within CAN messages, maintaining high efficiency without compromising performance~\cite{bnn3}. However, they focus only on payload data due to the inherent ease of binary encoding, whereas time-series characteristics and message IDs are also critical.
\section{System Model and Threat Formulation}
\begin{figure}
\centering\includegraphics[width=1\linewidth]{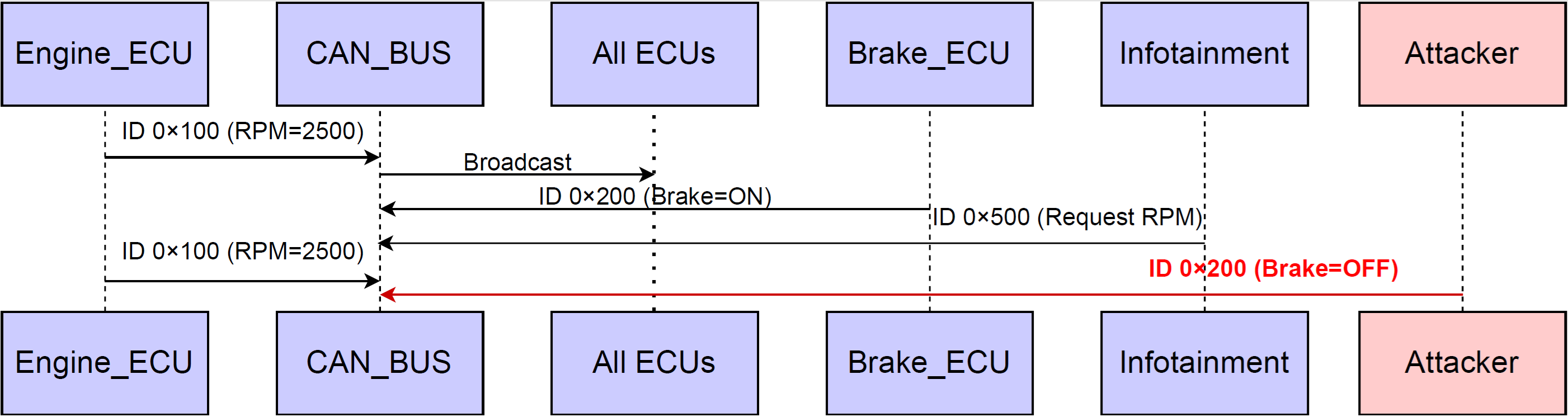}
    \caption{General Communication Diagram}
    \label{cancom}
\end{figure}
Figure~\ref{cancom} illustrates the in-vehicle network communication, encompassing both legitimate and malicious messages. ECUs are embedded controllers tasked with managing specific vehicle subsystems, while the CAN bus functions as the physical and logical backbone for in-vehicle communication, typically implemented using a twisted pair cable. The rest of this section briefly introduces the task formulation of in-vehicle intrusion detection and attack model.

\subsection{Task Formulation}
Let an edited CAN message at time $t$
be represented as:
\begin{equation}
    m_t = (ID_t, p_t, \Delta t)
\end{equation}
where $ID_t$ serves as an identifier, $p_t \in \{0,1\}^{64}$ is the binary payload, and $\Delta t$ is the time interval calculated based on the current timestamp and the previous from the same ID.

The anomaly detection can be defined as a binary classification task as~\eqref{eqb}.
\begin{equation}\label{eqb}
    f_{\text{binary}}(m_t) =
\begin{cases}
1 & \text{(attack) if } P(m_t \notin N) \geq \tau \\
0 & \text{(benign) otherwise}
\end{cases}
\end{equation}
where $N$ is the normal message distribution and $\tau$ is the detection threshold for safety-critical systems.

The classification of different types of network traffic can be formulated as a multi-class classification such as~\eqref{multi}.
\begin{equation}\label{multi}
    f_{\text{multi}}(m_t) = c \in {0, 1, \ldots, C}
\end{equation}
where 
$c = 0 \text{ (benign), } c \geq 1 \text{ (DoS, spoofing, \ldots)}.$

\subsection{Attack Model}
A typical CAN message for ECU communication is illustrated below, where the ID and Data (payload) are critical for security analysis. The broadcast nature of CAN allows messages to reach all ECUs without authentication, and payloads are transmitted as plain text without encryption. This vulnerability facilitates the initiation of various attacks, as demonstrated in the following examples.
\[\footnotesize
{\begin{array}{|c|c|c|c|c|c|c|c|}
    \text{SOF} & \text{ID} & \text{RTR} & \text{Control} & \text{Data} & \text{CRC} & \text{ACK} & \text{EOF} \\
    \textit{begin} & \textit{11/29b} & \textit{Remote} & \textit{4b} & \textit{0-64b} & \textit{15b} & \textit{1b} & \textit{end} \\
\end{array}}
\]
\begin{itemize}
    \item  Flooding: Flood bus with high-priority messages, with abnormal concentration of certain ID. ($\Delta t_{attack} \approx 0$)
    \item Fuzzing: Inject random IDs/payloads to crash ECUs ($ID \notin allowlist$ or $p_{attack} \sim U\{0,1\}^{64}$)
    \item Spoofing: Forge legitimate ID with malicious payload. ($p_{spoofing}\neq p_{benign}$)
\end{itemize}
As demonstrated, time, ID, and payload are critical for attack detection, and the fusion of these features enables cross-validation of anomalies.

\section{Proposed Method}
\begin{figure}
    \centering
\includegraphics[width=1\linewidth]{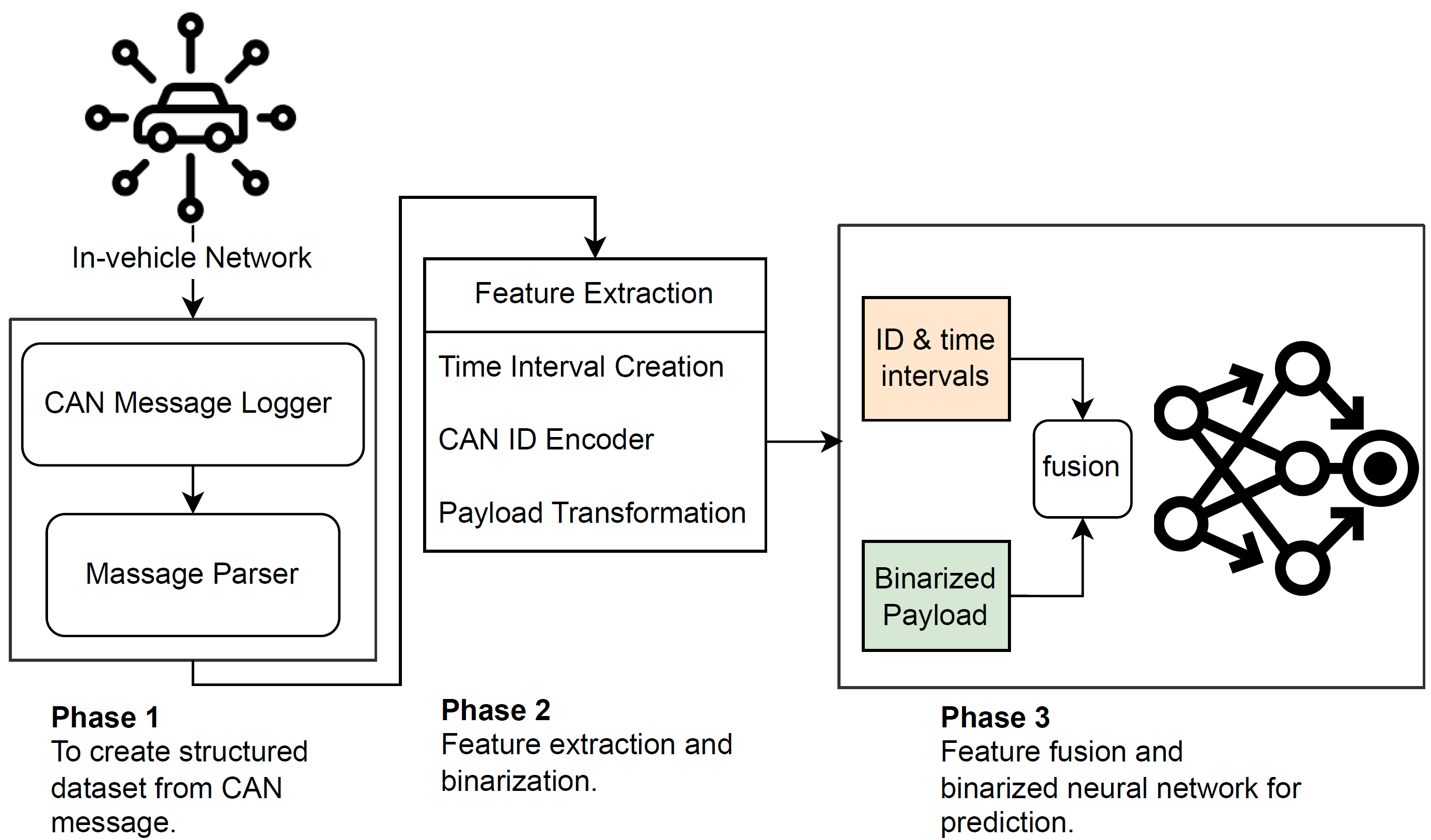}
    \caption{High-level architectural view of the proposed method.}
    \label{workflow}
\end{figure}
Figure \ref{workflow} presents a high-level overview of the proposed methods, including CAN traffic data logging and parsing, CAN-specific feature extraction and binarization, and feature fusion and BNN-based classification. 
    
The proposed intrusion detection method processes individual CAN messages by extracting three critical components: 
\begin{enumerate}
    \item Time Interval: Binarized into threshold-based 3 states
    \item Message ID: Encoded as a 6-bit binary vector
    \item Payload: Each byte converted to 8 bits (64 bits total)
\end{enumerate}
This results in a 73-dimensional feature vector per message (6 + 3 + 64). These binary features are subsequently processed through a simple yet effective BNN employing a Straight-Through Estimator (STE) gradient approximation for optimization, enabling efficient hardware deployment while preserving detection accuracy.
\subsection{Feature Data Binarization}
We design different data processing methods to transfer raw CAN logs to ML-ready tensors for BNN.

CAN IDs are converted into a binary representation using a fixed number of bits (6-8 bits), depending on the number of distinct IDs in the network environment. The output of this conversion is a list of integers representing the binary digits. For example, with an ID of 25 and a default of 6 bits, the result is ``011001''. 

For the timestamp, we first calculate the time intervals by determining the difference between the current and previous timestamps for the same CAN ID. These intervals are then converted into a 3-bit representation using two thresholds. The thresholds are established based on the visualization of different traffic types to identify an abnormal range (e.g., excessively short intervals may indicate DoS attacks, which generate an unusually high volume of traffic to disrupt service). The conversion logic is outlined below:
\begin{itemize}
    \item val $<$ THRES\_1: [0, 0, 0]
    \item THRES\_1 $\leq$ val $<$ THRES\_2: [0, 0, 1]
    \item val $\geq$ THRES\_2: [0, 1, 0]
\end{itemize}
This simplifies the continuous time intervals into discrete categories, making them manageable for BNN models.

For CAN bus payload data, we convert each hexadecimal character into a 4-bit format. Given a data sample in hexadecimal as 
[\text{Hex:} 05 , 28 , 84 , 66 , 6d , 00 , 00 , a2], the conversion from each byte to binary can be represented below.
\begin{small}
\begin{align*}
05 & \rightarrow 0000 , 0101 ;28  \rightarrow 0010 , 1000 \\
84 & \rightarrow 1000 , 0100 ;66 \rightarrow 0110 , 0110 \\
6d & \rightarrow 0110 , 1101 ;00  \rightarrow 0000 , 0000 \\
00 & \rightarrow 0000 , 0000 ;a2  \rightarrow 1010 , 0010 \\
\end{align*}
\end{small}

\subsection{Binarized Neural Network}
BNNs are a special type of DL models where weights and activations are constrained to 
\{-1, +1\}~\cite{2016binarized}, which makes them computationally efficient and suitable for resource-constrained environments (e.g., IoT or embedded systems). Below are the key mathematical operations in the proposed BNN.

\textbf{Dot Product Approximation.} After the input binarization, we use the edited feed forward layers to perform the dot product approximation. The dot products with binarized weights simplify into bit-wise operations as~\eqref{bit}.
\begin{equation}\label{bit}
    z = W \cdot X + b \quad \text{for} \quad W \in \{-1, +1\}
\end{equation}
It performs XNOR operations, followed by bit counts.

\textbf{Layer Activations of Forward Pass.} Each layer performs: 
\begin{equation}
z^{(l)}=BatchNorm(Sign(W^{(l)}\cdot a^{(l-1)} +b))
\end{equation}
where $W^{(l)}$ is the binarized weight matrix, $a^{(l-1)}$ is the activation from the previous layer, and $sign(\cdot)$ operation enforces binary activations after batch normalization.

The remaining components follow standard neural network configurations: Batch normalization rescales and centers inputs, promoting faster convergence. For final outputs, a sigmoid activation is used for binary classification to convert raw logits into probabilities, while softmax is employed for probabilistic interpretation in multi-class classification.

\begin{figure}
    \centering
    \includegraphics[width=0.6\linewidth]{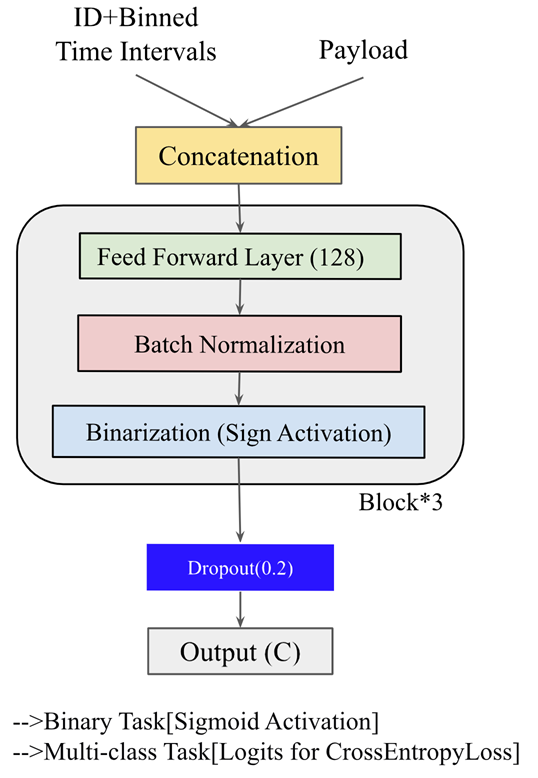}
    \caption{Binarized Neural Network used in the proposed method.}
    \label{fig:bnn}
\end{figure}

Figure~\ref{fig:bnn} displays the model architecture of the BNNs utilized in this study. The information fusion layer concatenates the processed payload data with binarized ID and time intervals. Each of the three fully connected layers comprises 128 neurons. Batch normalization layers are implemented following each linear layer to facilitate faster convergence and to mitigate internal covariate shift. The activation function  $sign(\cdot)$ is employed to binarized outputs, thereby mimicking the functionality of BNNs. Following the binarization step, dropout regularization is applied to reduce overfitting. Finally, the output layer is configured for either binary or multi-class classification, as required by the specific task. 

\section{Empirical Evaluation}
\label{sec: evaluation}
\subsection{Data Description and Experimental Settings}
This paper selects two realistic datasets for experiments.
CAN-IDS, collected in a real vehicle~\cite{canids}, serves as a benchmark for in-vehicle intrusion detection. It contains 49.5\% malicious and 50.5\% benign samples, distributed as follows:
\begin{itemize}
    \item Benign behavior: 2,268,488 samples ($50.5\%$)
    \item DoS attack: 645,754 samples ($14.4\%$)
    \item Fuzzy attack: 579,878 samples ($12.9\%$)
    \item Impersonation attack: 995,441 samples ($22.2\%$) 
\end{itemize}
The Car-Hacking dataset~\cite{carhacking} is derived from a real vehicle network with 14.07\% synthetic attack injections.
\begin{itemize}
    \item Benign behavior: 14,237,959 samples ($85.93\%$)
    \item Fuzzy attack: 491,847 samples ($2.97\%$)
    \item DoS attack: 587,521 samples ($3.55\%$)
    \item Spoofing\_RPM attack: 654,897 samples ($3.95\%$)
    \item Spoofing\_gear attack: 597,252 samples ($3.60\%$)
\end{itemize}

Stratified sampling is applied for training and validation, ensuring the splits preserve the original multi-class label distributions and reflect realistic traffic scenarios. 

The implementation and experiments utilize PyTorch. For binary classification, we employ BCEWithLogitsLoss, while for multi-class classification, CrossEntropyLoss is applied. We utilize the Adam optimizer with an initial learning rate of 0.001, incorporating a learning rate scheduler that reduces the rate by a factor of 0.1 if the validation loss plateaus for three consecutive epochs. To prevent exploding gradients, we apply gradient clipping with a maximum norm of 1.0. The training is set for 100 epochs, with early stopping triggered if no improvement in validation loss is observed for six epochs. 

The baseline DL models employed in this experiment consist of straightforward architectures utilizing only DNN, RNN, and LSTM layers in each. Each model is structured identically, comprising three layers with 64 units per layer. This design ensures a fair comparison against our BNN model.
\subsection{Experimental Results}
The experiments focus on two aspects: detection performance under binary and multi-class settings, specifically predicting whether traffic is malicious or classifying network traffic into multiple categories. The second aspect is model efficiency, which primarily addresses the size of the trained weights and response time.
\begin{table}[]
\caption{Performance Metrics on CAN-IDS Dataset (in \%)}
\label{canids}
\resizebox{\columnwidth}{!}{%
\begin{tabular}{lllllllll}
\hline
                    & \multicolumn{2}{l}{Binary   Classification}       &           &        & \multicolumn{3}{l}{Multi-class   Classification} &                                                                 \\
\hline
Model               & Accuracy                     & F1                 & Precision & Recall & Accuracy       & F1          & Precision         & Recall                                                          \\
DNN                 & 92.30                        & 91.93              & 95.47     & 88.64  & 92.49          & 93.48       & 95.03             & 92.48                                                           \\
RNN                 & 92.48                        & 92.10              & 95.97     & 88.52  & 92.37          & 93.42       & 94.75             & 92.52                                                           \\
LSTM                & 92.59                        & 92.24              & 95.74     & 88.98  & 92.43          & 93.45       & 94.86             & 92.52                                                           \\
BNN\_payload        & 92.39                        & 92.31              & 92.31     & 92.30  & 84.90          & 89.85       & 89.90             & 89.80                                                           \\
BNN\_full           & 92.97                        & 92.87              & 93.19     & 92.55  & 94.00          & 93.45       & 93.50             & 93.40                                                           \\
\hline
\multicolumn{2}{l}{DL-TSA   \cite{maliha2024unified}} &                    &           &        &                &             &                   & \begin{tabular}[c]{@{}l@{}}DoS 98.7\\ Fuzzy 69.9\end{tabular} \\
\multicolumn{2}{l}{AE+Attention   \cite{key_01}}     &   &           & 81.1/83.7  &                &             &       &    \\
\hline
\end{tabular}%
}
\end{table}

Table~\ref{canids} presents the performance metrics on CAN-IDS. As previous studies have used only payload data as input, we report results from two experiments: BNN\_payload (payload only) and BNN\_full (full data) for comprehensive comparison.

The BNN\_payload variant achieves accuracy (92.39\%) and F1 score (92.31\%) comparable to DNN, RNN, and LSTM in binary classification, albeit with slightly lower precision. However, it performs notably worse (accuracy ~85\%) on multi-class tasks compared to these models. This deficiency arises because using only payload data overlooks key temporal and structural features (such as ID and interval), which are critical for finer-grained classification. Conversely, BNN\_full achieves superior performance in binary classification, with the highest accuracy (92.97\%) and F1 score (92.87\%). It also demonstrates strong results in multi-class classification, attaining the best accuracy (94\%) and a favorable F1 score. These findings indicate that the BNN model's effectiveness significantly increases when input is not restricted to payload, as the full feature set provides richer contextual information.

Compared to other DL models in the literature, BNN remains competitive. DL-TSA~\cite{maliha2024unified} yields a higher detection rate for Dos attack but much lower performance on fuzzy attack. AE+Attention~\cite{key_01} was evaluated on a different data partition and demonstrated lower precision. These results suggest that increased model complexity does not guarantee improved performance, and such models do not surpass BNN.

\begin{table*}[]
\centering
\caption{Performance Metrics on Car-Hacking Dataset (in \%)}
\label{carhack}
\resizebox{0.9\textwidth}{!}{%
\begin{tabular}{llrllllllll}
\hline
& \multicolumn{5}{l}{Binary   Classification}   & \multicolumn{5}{l}{Multi-class   Classification}       \\
\hline
Model                                                               & Accuracy                      & \multicolumn{1}{l}{F1}                            & Precision                     & Recall                        & Confusion Matrix                                                    & Accuracy & F1                         & Precision & Recall                     & Confusion Matrix   \\
\hline
DNN                                                                 & 100.0    & 99.99                                            & 100.0      & 99.99   & \begin{tabular}[c]{@{}l@{}}3806141 11\\ 48  583387\end{tabular}     & 100.0       & 99.99  & 100.0         & 99.99  & \begin{tabular}[c]{@{}l@{}}3806136       0        16       0       0\\             0  147311         0       0       0\\            53       0    123063       0       1\\             0       0         0  149271       0\\             0       0         0       0  163736\end{tabular} \\
RNN  & 100.0  & 100.0  & 100.0       & 99.99    & \begin{tabular}[c]{@{}l@{}}3806137 15\\ 46  583389\end{tabular}     & 100.0 & 100.0 & 100.0 & 100.0 & \begin{tabular}[c]{@{}l@{}}3806141       0        11       0       0\\             0  147311         0       0       0\\            33       0    123084       0       0\\             0       0         0  149271       0\\             0       0         0       0  163736\end{tabular} \\
LSTM & 100.0   & 100.0 & 100.0 & 100.0 & \begin{tabular}[c]{@{}l@{}}3806143 9\\ 46  583389\end{tabular}      & 100.0   & 100.0 & 100.0 & 100.0  & \begin{tabular}[c]{@{}l@{}}3805978       0        22       0       0\\             0  146487         0       0       0\\            26       0    123304       0       1\\             0       0         0  149642       0\\             0       0         0       0  164127\end{tabular} \\
BNN\_payload                                                        & 99.84                        & 99.41  & 98.85                        & 99.97                        & \begin{tabular}[c]{@{}l@{}}3799942 6734\\  295  582616\end{tabular} & 99.84   & 99.52                     & 99.13    & 99.93                     & \begin{tabular}[c]{@{}l@{}}3800023    6609        44       0       0\\             0  146761         0       0       0\\           219       0    122779       2       0\\             0       0         0  149243       0\\             0       0         0       0  163907\end{tabular} \\
BNN\_full                                                           & 100.0    & 99.99   & 100.0       & 100.0    & \begin{tabular}[c]{@{}l@{}}3806143 9\\ 48 583389\end{tabular}       & 100.0    & 100.0               & 100.0    & 100.0     & \begin{tabular}[c]{@{}l@{}}3806136       0        16       0       0\\             0  147311         0       0       0\\            30       0    123086       0       1\\             0       0         0  149271       0\\             0       0         0       0  163736\end{tabular} \\
\hline
DNN\cite{okokpujie2021anomaly}                                      &  &   & 95  & 95  &  &          &  &           & &             \\
LeNet\cite{s21144736}                             & 98.1  & 97.83 & 98.14 &  98.04 &   &          &                            &           &                            &  \\
LSTM+CNN\cite{10556797} & 99.595 &99.58                      & 99.58  & 99.57  &                                                                     &          &                            &           &                            & \\     
\hline       
\end{tabular}%
}
\end{table*}
Table~\ref{carhack} reports the evaluation metrics on Car-Hacking dataset. As most metrics approach unity, and many figures are identical after rounding, confusion matrices are also listed to enable clear differentiation; nonetheless, the differences are negligible in practical terms.

Consistent with earlier observations, employing only payload data leads to a marked reduction in performance across all evaluation metrics, as seen in the BNN\_payload results (e.g., binary accuracy: 99.84\%, multi-class F1: 99.52\%). In contrast, utilization of the full dataset restores the competitiveness with DL baselines, as demonstrated by BNN\_full, which achieves excellent performance metrics (all near 100\%).

When compared with RNN and LSTM, which demonstrate superior results, BNN occasionally underperforms. This can be attributed to the fact that RNN and LSTM are specifically designed to capture temporal dependencies and sequence patterns; for data involving time-ordered events such as CAN messages, LSTM models these relationships more effectively. In contrast, BNN offers the advantages of probabilistic uncertainty estimation and robustness, but its core (feedforward/MLP) structure lacks the inherent ability to model temporal dependencies.

Nonetheless, BNN proves highly effective for outlier detection and benefits from a simpler architecture, resulting in greater computational and memory efficiency, which is an important consideration for embedded environments with tight resource constraints. Even when the full feature set increases input dimensionality, BNN maintains competitive classification performance while still providing efficiency advantages over RNN and LSTM in certain embedded scenarios.

\begin{table}[]
\caption{Comparison of Model Efficiency: Size and Response Times Across Architectures}
\label{tabeff}
\resizebox{\columnwidth}{!}{%
\begin{tabular}{llll}
\hline
Model & Model Size                       & Response Time (CUDA) & Response Time (CPU) \\
\hline
DNN   & 39.91 kB                         & 0.0019 ms            & 0.0087 ms           \\
RNN   & 89.14 kB  & 0.0028 ms            & 0.0214 ms           \\
LSTM  & 347.33 kB & 0.0046 ms            & 0.0386 ms           \\
BNN   & 156.0 kB                         & 0.0018 ms            & 0.0061 ms\\
\hline  
\end{tabular}%
}
\end{table}

Table~\ref{tabeff} presents the size of trained weights and response times under both CPU and GPU conditions. Although BNN contains more parameters than DNN and RNN, owing to the increased number of units (128 per layer) and additional operations, it remains lightweight overall. Furthermore, BNN achieves the fastest response times in both scenarios, while RNN and LSTM are considerably slower, particularly on CPU.

In environments with highly constrained resources (RAM, CPU, power), BNN is the most suitable choice for intrusion detection. For applications requiring uncertainty quantification, BNN is also preferable to deterministic DNN. Where maximum raw classification performance is the priority and resources are sufficient, LSTM offers the best results due to its proficiency in modeling temporal patterns, but this comes at increased resource cost. Nevertheless, across all scenarios, BNN provides robustness and enhanced interpretability.

\textbf{Future Enhancement.} BNN performance could be further enhanced while maintaining its lightweight and efficient characteristics through novel optimization techniques and training strategies. BNext~\cite{guo2023towards} introduced an optimization pipeline integrating a robust BNN, a teacher-student knowledge complexity metric, and a multi-round knowledge distillation technique known as consecutive knowledge distillation, which improved performance in computer vision tasks. SGDAT~\cite{sgdat} addressed the limitations of standard SGD in BNNs by optimizing binary weights separately with an adaptive threshold strategy, thereby enhancing training stability and accuracy. Another similar approach involves using a probabilistic optimizer that weights signs with Bernoulli distributions of accumulated gradients~\cite{probabilistic}. Lastly, Inverse Binary optimization (IBO)~\cite{park2025inverse} proposed a surrogate function based on CNNs to enhance binary network training, improving accuracy by more effectively approximating the binary optimization landscape.

The proposed BNN-based IDS offers notable performance and computational advantages, yet several challenges hinder its deployment in diverse vehicular environments. A key issue is hardware compatibility, as not all in-vehicle ECUs or micro-controllers support efficient execution of BNN bit-level operations. In addition, integrating BNN detection techniques with existing in-vehicle networks may require tailored adaptation or modular deployment strategies. Future research should focus on co-design with hardware and system integrators to address these challenges, facilitating practical adoption across a wider range of vehicular platforms.

\section{Conclusion}
This paper addresses the critical security vulnerabilities of the CAN protocol and the practical deployment challenges associated with complex ML/DL-based IDSs by introducing a novel, lightweight intrusion detection framework based on BNNs, specifically designed for in-vehicle environments. The experimental results consistently demonstrate that the proposed BNN framework achieves competitive performance and exceptional efficiency. Although LSTM models exhibited marginally superior performance in one experimental context, the difference was minimal. Importantly, the BNN framework is significantly more suitable for deployment in resource-constrained embedded environments, thereby aligning seamlessly with the real-time safety requirements of CAN bus applications. This work highlights the potential of lightweight BNNs to provide effective and practical security solutions for vehicular networks.

\bibliographystyle{IEEEtran}
\bibliography{ieeeref}
\end{document}